\documentclass{aa} % AA vers. 8.2, LaTeX class for Astronomy & Astrophysics
\usepackage{amsmath,amssymb}%, amsthm}
\usepackage{mathtools}
\usepackage{color}
\usepackage{soul}
\usepackage{natbib}
\usepackage{hyperref}

\DeclarePairedDelimiter\ket{\lvert}{\rangle}
\DeclarePairedDelimiterX\braket[2]{\langle}{\rangle}{#1 \delimsize\vert #2}
\definecolor{orange}{RGB}{255,127,0}
\begin{document}
\title{
{Fast and accurate approximation of the
angle-averaged redistribution function} 
{for polarized radiation}}
\author{A. Paganini 
        \inst{1}
        \and
        B. Hashemi
        \inst{2, 3}
        \and
        E. Alsina Ballester
        \inst{4}
        \and
        L. Belluzzi
        \inst{4, 5}}
\institute{School of Mathematics and Actuarial Science, 
           University of Leicester,
           LE1 7RH Leicester, 
           United Kingdom
           \and 
           Department of Mathematics, 
           Shiraz University of Technology, 
           Modarres BLVD, 
           Shiraz 71555-313, 
           Iran
           \and 
           School of Mathematics, 
           Institute for Research in Fundamental Sciences (IPM), 
           P.O. Box: 19395-5746, Tehran, Iran
           \and
           Istituto Ricerche Solari Locarno,
            6605 Locarno-Monti,
            Switzerland
            \and
            Leibniz-Institut f\"{u}r Sonnenphysik (KIS),
            79104 Freiburg, 
            Germany}           
 \abstract
 {{Modeling spectral line profiles taking frequency redistribution effects into
 account is a notoriously challenging problem from the computational point of view, especially when
 polarization phenomena (atomic polarization and polarized radiation) are taken into account. 
Frequency redistribution effects are conveniently described through the redistribution function formalism, 
and the angle-averaged approximation is often introduced to simplify the problem. Even in this case, the evaluation 
of the emission coefficient for polarized radiation 
remains computationally costly, especially when magnetic fields are present or complex atomic models are considered.}} 
{{We aim to develop}
an efficient algorithm
to 
{numerically evaluate}
the angle-averaged
redistribution function 
{for polarized radiation.}} 
{{The proposed approach
is based on {a} low-rank approximation {via}}
trivariate polynomials
whose univariate components are represented in the
Chebyshev basis.}
{{The {resulting algorithm is} significantly
 faster than standard quadrature-based schemes
for any target accuracy in the range {$[10^{-6},10^{-2}]$}}.}
{}
 \date{}    
%%%%%%%%%%%%%%%%%%%%%%%%%%%%%%%%%%%%%%%%%%%%%%%%%%%%%%%%%%
%%%%%%%%%%%%%%%%%%%%%%%%%%%%%%%%%%%%%%%%%%%%%%%%%%%%%%%%%%
%
\titlerunning{Numerical Approximation of $R_{\mathrm{II-AA}}$}
\authorrunning{Paganini et al.}

\maketitle  

\section{{Introduction}}
\label{sec:intro}
Synthesizing spectral line profiles through radiative transfer (RT) 
calculations, out of local thermodynamical equilibrium {conditions,} 
{is} a key problem in current solar 
and stellar physics research. 
{In solar physics, particular attention is nowadays paid to the RT modeling of 
the polarization profiles of strong resonance lines {because} they encode {valuable} 
information on the magnetic properties of two atmospheric layers of high 
scientific interest, namely the chromosphere and transition region.
Unfortunately, the well-known Zeeman effect turns out to be of limited utility 
for investigating the weak and tangled magnetic fields that are generally 
present in these hot regions, and alternative diagnostic methods are 
therefore being developed.
Chief among them are those that exploit the combined action of scattering 
polarization (i.e., polarization produced by the scattering of anisotropic 
radiation), along with its modification  
{due to} 
the presence of a magnetic 
field (Hanle effect), and the Zeeman effect \citep{TrujilloBueno14}.
The information on the magnetism of the upper chromosphere that was made available from 
the data acquired through the recent Chromospheric Lyman-Alpha 
SpectroPolarimeter (CLASP) experiment is just one example of the success of 
such approaches \citep{Kano+17,TrujilloBueno+18}.}

A major difficulty in modeling strong resonance lines is the need to account 
for frequency correlations between incoming and outgoing photons in scattering 
processes. 
{Taking partial frequency redistribution (PRD) effects into account is} 
{necessary in order to correctly model}
{the wings of the intensity profiles in strong 
spectral lines, and is crucial for reproducing the large scattering polarization signals that are observed in 
the wings of such lines.}
{Modeling PRD effects is notoriously difficult, especially from the 
computational point of view, representing} 
{a true challenge (both theoretically and numerically) 
when scattering polarization and the Hanle and Zeeman effects are taken into 
account.}

{A convenient way to describe PRD phenomena is through the redistribution function formalism.}
{This formalism was initially developed for the unpolarized case \citep[e.g.,][]{Hummer62} and was subsequently 
generalized to the case in which polarization phenomena are taken into account.}
\subsection{{The unpolarized case}}
{Neglecting} polarization effects, the scattering contribution 
to the line emission coefficient, at reduced frequency $u$ and for direction 
$\vec{\Omega}$, is given by
\begin{align}
	\label{Eq:emis}
 	& \varepsilon_I(u,\vec{\Omega}) = k_L \nonumber \\
 	& \, \times \int_{-\infty}^\infty {\rm d} u^\prime
 	\oint \frac{{\rm d} \vec{\Omega}^\prime}{4 \pi}
	R(u^\prime,\vec{\Omega}^\prime,u,\vec{\Omega}) \, 
	I(u^\prime,\vec{\Omega}^\prime) \, ,
\end{align}
where $R(u^\prime,\vec{\Omega}^\prime,u,\vec{\Omega})$ is the redistribution 
function, $I(u^\prime,\vec{\Omega}^\prime)$ is the intensity of the incoming 
radiation, and $k_L$ is the frequency-integrated absorption coefficient.
Throughout this work, we follow the convention according to which the primed 
quantities refer to the incident radiation and the unprimed ones to the 
scattered radiation.

Throughout this work, we focus only on the redistribution function 
characterizing scattering processes that are coherent (in frequency) in the 
atomic rest frame. %,
Following the terminology introduced by \citet{Hummer62}, this redistribution 
function is referred to as $R_{\mathrm{II}}$. 
In the atomic reference frame, its frequency and angular dependencies are 
completely decoupled, but in the observer's frame the Doppler effect 
introduces a complex coupling of frequencies and angles, making the evaluation 
of $R_{\mathrm{II}}$ and of the integrals of Eq.~(\ref{Eq:emis}) very 
demanding from the computational point of view.

\subsubsection{{The angle-averaged approximation}}
{In order to make the problem computationally simpler, \citep{ReesSaliba82} 
proposed the so-called {angle-averaged} approximation,}
which allows decoupling the frequency and angular dependencies also in the 
observer's frame. 
Under this assumption, $R_{\mathrm{II}}$ can be written as
\begin{equation}
	\label{Eq:redisfac}
	R_{\mathrm{II-AA}}(u^\prime,\vec{\Omega}^\prime,u,\vec{\Omega}) = 
	\mathcal{R}_{\mathrm{II-AA}}(u^\prime,u) \, 
	\mathcal{P}(\vec{\Omega}^\prime,\vec{\Omega}) \, .
\end{equation}
The quantity $\mathcal{P}$ is the so-called angular phase function. Its 
{explicit expression} is not relevant 
for the following discussion. 
The explicit form of the frequency-dependent part, {for the 
simplest case of a two-level atom with an infinitely sharp lower level}, is
\begin{align}
	& \mathcal{R}_{\mathrm{II-AA}} (u^\prime,u) = 
	\frac{\Gamma_R}{\Gamma_R + \Gamma_I + \Gamma_E} \, 
	\nonumber \\
	& \qquad \; 
	\times \frac{1}{2 \pi} \int_0^{\pi} {\rm d} \Theta
	\exp{\left[ -\left( \frac{u^\prime-u}{2 \sin{\Theta/2}} \right)^2 \right]} 
	\nonumber \\ \, 
	& \qquad \qquad \; \times 
	H \left(\frac{a}{\cos{\Theta/2}},\frac{u+u^\prime}{2 \cos{\Theta/2}}\right)
	\, ,
	\label{Eq:RII-AA_unpol}
\end{align}
where $\Gamma_R$, $\Gamma_I$, and $\Gamma_E$ are the broadening constants due 
to radiative decays, inelastic de-exciting collisions, and elastic collisions, 
respectively (these quantities only depend on the spatial point), $\Theta$ is 
the angle between the incoming and scattered direction (scattering angle), 
$H$ is the Voigt function, and $a$ is the damping constant, which 
depends only on the spatial point.

{It must be stressed that the angle-averaged approximation introduces considerable 
inaccuracies in the calculations of the emission coefficient.} {Thanks to  
increases in computational power over the last decades, this 
approximation has been progressively abandoned in favor of  
{angle dependent} PRD calculations,} {which allow for} {more reliable quantitative 
comparisons between synthetic and observed intensity profiles.}

\subsection{{The polarized case}}
We now consider the polarized case, accounting for scattering polarization 
and the Hanle and Zeeman effects.
We still refer to the simple case of a two-level atom, here also in the 
presence of a magnetic field. {However}, the following discussion also holds
for more complex atomic models, such as multilevel atoms, multiterm atoms, or 
atoms with hyperfine structure.
The polarization properties of the radiation field are commonly described 
through the four Stokes parameters, $I$, $Q$, $U$, and $V$ (where $I$ is the
usual intensity).   
In analogy with Eq.~(\ref{Eq:emis}), the emission coefficient in the four
Stokes parameters is given by
\begin{align}
	& \varepsilon_i(u,\vec{\Omega}) = k_L \nonumber \\
	& \quad \times \int_{-\infty}^{\infty} \!\! {\rm d} u^\prime \!
	\oint \frac{{\rm d} \vec{\Omega}^\prime}{4 \pi} \sum_{j=0}^3 
	\left[ R(u^\prime,\vec{\Omega}^\prime,u,\vec{\Omega}) \right]_{ij} 
	I_j(u^\prime,\vec{\Omega}^\prime) \, , 
\label{eq:emisPolar}
\end{align}
{where the indices $i$ and $j$ can take values $0,1,2,$ and 3, standing for 
Stokes $I$, $Q$, $U$, and $V$, respectively, and $[R]_{ij}$ is a $4 \times 4$
matrix that generalizes the concept of redistribution functions to the 
polarized case.}

Again, we focus on the redistribution matrix characterizing scattering 
processes that are coherent in the reference frame of the atom,  
$\bigl[R_{\mathrm{II}}\bigr]_{ij}$, 
accounting for Doppler redistribution in the 
reference frame of the observer. 
Its expression is significantly more complex than in the unpolarized case
because, in order to model scattering polarization, it is necessary 
to provide a complete description of the atomic system, specifying the 
population of each magnetic sublevel as well as the coherence that may be 
present between pairs of sublevels.
The inclusion of the latter physical ingredient is responsible for the 
appearance in the redistribution matrix of additional terms that, 
instead of the Voigt profile $H$, involve the 
associated dispersion profile $L$ {\citep[e.g.,][]{BLandiLandolfi04}}. 
Moreover, when a magnetic field is present, the redistribution matrix is given by
a linear combination of various terms, 
each being associated to a particular scattering channel 
$\ket{\ell} \rightarrow \ket{u} \rightarrow \ket{\ell'}$, where 
$\ket{\ell}$ and $\ket{\ell'}$ indicate the initial and final lower magnetic 
sublevels involved in the scattering process, and $\ket{u}$ the intermediate 
upper magnetic sublevel. 
The various terms are shifted in frequency with respect to each other due to 
the energy shifts induced in the presence of a magnetic 
field.\footnote{When a multilevel atom, a multiterm atom, or an atom with hyperfine 
structure is considered, the lower sublevels $\ket{\ell}$ and $\ket{\ell'}$ can 
pertain to different fine structure or hyperfine structure levels (Raman 
scattering). In this case, the redistribution matrix is given by a linear 
combination of various terms also in the absence of magnetic fields.}

{\subsubsection{{The angle-averaged approximation}}}
{Even with the computational resources that are nowadays commonly available, }
dealing with the general, angle-dependent expression of 
$\bigl[R_{\mathrm{II}}\bigr]_{ij}$ is a formidable task.   
{For this reason}, it is customary to introduce, in full 
analogy with the unpolarized case, the angle-averaged assumption. 
As shown in detail in Appendix \ref{app:A} for the case of a two-level atom 
in the presence of a magnetic field, the frequency-dependent part of
$\bigl[R_{\mathrm{II-AA}}\bigr]_{i j}$ is given by a linear combination of 
functions of the form
\begin{align}
	f(x,y,a) = & \frac{1}{\pi} \int_0^{\pi/2}  {\rm d} \gamma 
	\exp{\left( \frac{-x^2}{\sin^2 \gamma} \right)} \, 
	H \! \left( \frac{a}{\cos{\gamma}}, \frac{y}{\cos{\gamma}} \right) , 
    \label{eq:ffun} \\
	h(x,y,a) = & \frac{1}{\pi} \int_0^{\pi/2} {\rm d} \gamma 
	\exp{\left( \frac{-x^2}{\sin^2 \gamma} \right)} \, 
	L \! \left( \frac{a}{\cos{\gamma}}, \frac{y}{\cos{\gamma}} \right) \, ,
\label{eq:hfun}
\end{align}
where we have introduced the angle $\gamma = \Theta/2$, and where $H$ and $L$ 
are the Voigt and the associated dispersion profiles, respectively. 
The arguments $x$ and $y$, which take into account the magnetic splitting of 
the Zeeman sublevels, are given by
\begin{align}
	& x = \frac{u^\prime - u -\Delta_{\ell \ell^\prime}}{2} \, , \\
	& y = \frac{u^\prime + u +\Delta_{u \ell} + \Delta_{u \ell^\prime}}{2} \, , 
	\label{eq:arguments}
\end{align} 
where $\Delta_{\ell \ell'}$ is the frequency splitting (in Doppler width units) 
between lower sublevels $\ell$ and $\ell'$, while $\Delta_{u \ell}$ is the 
frequency shift (in Doppler width units) of the Zeeman transition between
sublevels $u$ and $\ell$ with respect to the line-center frequency.
We note that, when $\Delta_{u \ell}$ and $\Delta_{\ell \ell^\prime}$ are zero, 
the $f$ function is equivalent to the integral appearing in 
Eq.~\eqref{Eq:RII-AA_unpol}, inclusive of the $1/(2\pi)$ factor. 

{Like in the unpolarized case,}{the angle-averaged assumption 
represents a strong 
approximation, {as it} smoothens geometrical aspects of the problem that 
may have a significant impact on polarization.
Nonetheless,} {there is still} {high interest in modeling scattering 
polarization under this approximation. 
The main reason is that the computational cost for carrying out detailed 
angle-dependent calculations in the presence of polarization phenomena is 
still prohibitive except when taking the simplest modeling of the 
solar atmosphere \citep[see][]{Sampoorna+19} or when introducing other 
simplifying assumptions such as cylindrical symmetry 
\citep[see][]{delPinoAleman+20}.} 
{It is also for this reason that, to date, few quantitative analyses
of the impact of the angle-averaged assumption on the modeling of scattering polarization
have been carried out.}  
{Theoretical investigations performed in isothermal atmospheric models
\citep[see][]{Sampoorna+17,NagendraSampoorna11} have shown that this 
approximation can give rise to significant inaccuracies, although} {mainly}
{in the core of the scattering polarization signals.}

{Nevertheless, under the angle-averaged approximation it is still possible 
to conduct investigations of scientific interest, especially regarding the 
modeling of the large polarization signals produced by coherent scattering 
processes in the wings of strong resonance lines. 
Although approximate, an angle-averaged PRD approach contains the relevant 
physics (coherent scattering), and allows modeling such signals, which would 
be completely lost in a complete frequency redistribution (CRD) approach.
For instance, by making an angle-averaged PRD modeling of the H~{\sc{i}}
Lyman-$\alpha$ line, \cite{Belluzzi+12} predicted that linear polarization
signals of large amplitude should be found in the wings of this line. 
This theoretical result was subsequently confirmed by the observations carried 
out by the CLASP sounding rocket experiment \citep{Kano+17}.} 
{Moreover, the recent discovery that the wing scattering polarization signals 
in a variety of spectral lines are sensitive to magnetic fields through 
magneto-optical effects \citep{delPinoAleman+16,Alsina+16,AlsinaBallester+18,AlsinaBallester+19} has 
further awakened the interest in modeling 
their polarization profiles taking PRD effects into account, 
also in scenarios that can only be feasibly considered by making the angle-averaged approximation. }

{We finally observe that, at present, PRD scattering 
polarization calculations can only be performed in one-dimensional (1D) models of the 
solar atmosphere.
The angle-averaged approximation would not be justified in 3D  
because it would cancel out important geometrical effects, thus negating the effort of 
making a full 3D modeling. On the other hand, when accepting the simplifications of a 1D modeling it still 
represents a good compromise between accounting for the relevant physics of 
the problem and reducing the computational requirements of a PRD calculation.}
\\

\subsubsection{{The computational challenge}}
{Even under the angle-averaged assumption,} calculating the
emission coefficient {in the four Stokes parameters} at a given 
spatial point, frequency, and direction is a computationally demanding task 
that requires many evaluations of the $\bigl[R_{\mathrm{II-AA}}\bigr]_{ij}$ 
redistribution matrix. 
{As noted above, the latter is generally composed of many different  
terms containing the functions $f$ and $h$ for shifted values of $x$ and $y$. 
As an example, in order to model the Na~{\sc i} doublet at 589~nm, a two-term 
atom with hyperfine structure must be considered. The redistribution matrix 
for this atomic model, in the presence of magnetic fields, contains on the 
order of 100 distinct terms.} 
{Moreover, at each iterative step of the numerical solution
of a standard RT problem, the emission coefficient must 
be evaluated a considerable amount of times: at every spatial point in the 
considered model atmosphere and at all frequencies and propagation directions 
of the chosen frequency and angular grids of the problem.\footnote{Typical
frequency and angular grids contain roughly 100 frequency points and 100 
directions, respectively. Standard 1D semi-empirical models 
of the solar atmosphere contain roughly 100 spatial points (heights), while 
three-dimensional (3D) models obtained from MHD simulations may easily contain 
$500^3 = 1.25\cdot 10^{8}$ points.}} 
This clearly highlights the importance of developing faster methods for calculating
$\bigl[R_{\mathrm{II-AA}}\bigr]_{ij}$ and the integrals appearing in 
Eq.~(\ref{eq:emisPolar}), {without compromising the accuracy of the 
calculations beyond the level that can be achieved under 
the angle-averaged approximation}. 

A direct approximation of the functions $f$ and $h$
using quadrature rules is challenging because the integrands depend strongly on 
the values of $x$ and $y$. In particular, the integrands exhibit a steep decay 
to zero both for $\gamma \rightarrow 0$ when $0 < |x| \ll 1$, and for 
$\gamma \rightarrow \pi/2$ when $|y| \ll 1$. 
Therefore, an accurate approximation of these functions may require numerous 
quadrature points. Devising a strategy that, for each pair $(x,y)$, selects
a quadrature rule by balancing accuracy and computational cost is technical 
and tedious, and may need case-by-case adjustments.

To tackle this issue, approximations to quickly
evaluate the function $f$ have been proposed in the past 
\citep[e.g.,][]{Adams+71,Gouttebroze86,Uitenbroek89}. 
{Unfortunately}, these techniques cannot be directly extended
to the function $h$.

\subsection{Scope of the work}
The high scientific interest in modeling scattering polarization, taking PRD 
effects into account (even under the angle-averaged assumption) motivates the 
search for algorithms {through which} the functions appearing 
in Eqs.~\eqref{eq:ffun} and \eqref{eq:hfun} {can be evaluated 
at a lower computational cost than with existing competing strategies}. 
In this work, we propose a method based on Chebyshev polynomials to approximate
these functions. 
This method allows for a substantial speed-up with respect to quadrature-based 
approaches while keeping the error well below those originating from other 
approximations introduced in the problem (e.g., the angle-averaged 
assumption). 

This article is organized as follows.  
In Sects.~\ref{sec:approximation} and \ref{sec:FVoigt}, we describe and 
validate the new approximation algorithm for the functions $f$ and $h$, 
respectively. 
In Sect.~\ref{sec:application} we test its performance in a physically 
relevant application.  
Conclusions and perspectives are presented in Sect.~\ref{sec:concl}.

\section{Fast and accurate approximation of 
{$f$}}
\label{sec:approximation} 

{In this work we present a novel approach that involves replacing the
aforementioned functions $f$ and $h$, which have dependencies on $u$,
$u^\prime$, and $a$, by low-rank approximations in terms of a trivariate
polynomial. Its univariate components are represented in the Chebyshev basis.
Such polynomials can be constructed and stored easily using the
\textsc{Matlab}-software package \texttt{Chebfun}. 
{The} \textsc{Matlab}{-code used in the following sections is available at
 \cite{Zenodo}. Using} \textsc{Matlab}{'s product {Coder},\footnote{More
 information about \textsc{Matlab} Coder is available on \textsc{Matlab}'s
 website \url{https://www.mathworks.com/} 
.} the resulting approximations can
be exported as \texttt{C} or \texttt{C++} code to be used in existing
software for RT calculations.} 
}
In {the following section we focus our attention on the function $f$.}

\subsection{Restriction to a bounded domain}
The first step in constructing an approximation of $f$ is to identify the 
domain where the approximation needs to be 
accurate. 
Firstly, we observe that the integrand in Eq.~\eqref{eq:ffun} is an even 
function with respect to both $x$ and $y$. Therefore, we can restrict our 
considerations to the positive quadrant $x,y\geq 0$.

\begin{figure}[htb!]
\centering
\includegraphics[width=1\linewidth]{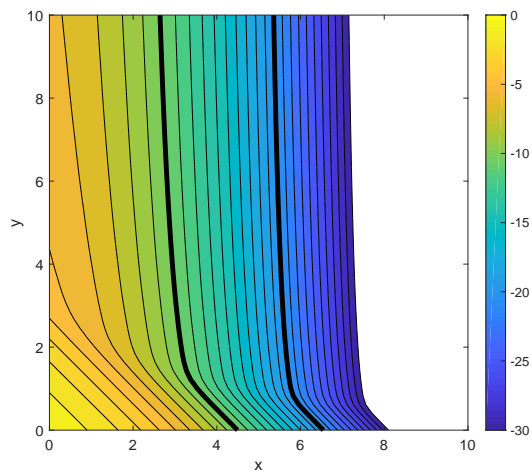}
\caption{Plot of $\log_{10}f(x,y,a)$ for $a=10^{-3}$ and $x,y\in[0,10]$.
The two bold lines indicate the $f=10^{-10}$ and the $f=10^{-20}$ contours.
Image values smaller than $10^{-30}$ are not plotted.
We observe a rapid decay in the $x$ direction. The qualitative behavior of
$\log_{10}f$ is similar for $a\in[10^{-5},10^{-1}]$.}
\label{fig:realRII}
\end{figure}

Secondly, $f(x,y,a)$ exhibits a super exponential decay in the variable $x$ 
and is smaller than $10^{-16}$ for $x>6$ (see Fig.~\ref{fig:realRII}).
Therefore, we can restrict our considerations to the interval $x\in[0, 6]$.

Finally, we do not observe any particularly notable behavior in the 
dependence on $y$ or $a$. We decide to consider the interval $y\in[0,10]$
and $a\in[10^{-5},10^{-1}]$ because these are the regimes that are typically 
considered in most applications.
We note that considering a larger interval for $y$ or $a$ presents no particular 
difficulty beyond a modest increase in computational cost.

\subsection{Interpolation with {Chebfun}}
Having identified the interpolation domain 
$[0, 6] \times [0, 10] \times [10^{-5},10^{-1}]$, we can proceed to the 
construction of a trivariate {polynomial} for approximating the 
function given by Eq.~\eqref{eq:ffun}. 
Since $f$ exhibits a fast decay (but remains positive), it is convenient to 
interpolate $\log f$ instead to better control the relative error.
Additionally, to have {equidimensional} ratios of the 
interpolation domain, we replace the variable $a$ with its base-10 logarithm 
$b$.
Therefore, we want to construct a trivariate polynomial $p$ such that
\begin{equation}
	\label{eq:approximant}
	p(x,y,b) \approx \log f(x,y,10^b) \, ,
\end{equation}
where $\log f$ denotes the natural logarithm of $f$.

Constructing a trivariate interpolant that is highly accurate is notoriously 
difficult. In this work, we use \texttt{Chebfun3} \citep{Chebfun3}, a component of the 
\textsc{Matlab}-software package \texttt{Chebfun} \citep{driscoll2014chebfun}. 
All \texttt{Chebfun3} needs to construct the interpolant $p$ 
is a \textsc{Matlab}-function that, for a triplet $(x,y,b)$, returns the value 
$\log f(x,y,10^b)$. {The details of computing $\log f$ are discussed in 
Sect.~\ref{sec:accurateintegration}.} 
Then, the interpolant $p$ can be computed with the following code
\begin{verbatim}
BD = [0,6,0,10,-5,-1];
p = Chebfun3(logf, BD, `eps', 1e-11).
\end{verbatim}
The first line of this code specifies the interpolation domain boundary, 
whereas the second line constructs the interpolant $p$. 
The option \texttt{`eps'} specifies the desired target accuracy. It is 
important to stress that setting \texttt{`eps'} to $10^{-11}$ does not
guarantee that $p$ has 10 digits of accuracy. 
Computing errors of trivariate interpolants is computationally expensive (as 
it requires evaluating $p$ over the whole interpolation domain).
Therefore, the software \texttt{Chebfun3} uses some heuristics to determine 
the accuracy of $p$.

The function $p$ returned by {Chebfun3} represents a trivariate 
polynomial in the following continuous analog of the Tucker 
decomposition of discrete tensors \cite[subsec. 12.5]{Golub13}
\begin{equation}
	\label{eq:Chebfun3}
	p(x,y,b) = \sum_{i=1}^{r_1} \sum_{j=1}^{r_2} \sum_{k=1}^{r_3} C_T(i,j,k) 
	c_i(x) s_j(y) t_k(b) \, ,
\end{equation}
where $C_T\in\mathbb{R}^{r_1\times r_2\times r_3}$ is the so-called core 
tensor, and $c_i(x)$, $s_i(y)$, and $t_k(b)$ are univariate polynomials. 

To construct (\ref{eq:Chebfun3}), Chebfun3 exploits low-rank compression of $f$ 
via multivariate adaptive cross approximation, which is an iterative 
application of a multivariate extension of Gaussian elimination with complete 
pivoting.
The trilinear rank $(r_1, r_2, r_3)$ as well as the degree of each set of 
polynomials $c_i, s_j, t_k$ are all chosen adaptively by the algorithm.
We refer to \cite{Chebfun3} for more details.

\subsection{Accurate evaluation of $f$ via integration}
\label{sec:accurateintegration}
To construct the interpolant $p$, we need an algorithm that evaluates
the function $\log f$, and thus $f$, to high accuracy.
This can be done using a Gauss quadrature formula.  %ASK ALBERTO

\begin{figure}[htb!]
\centering
\includegraphics[width=0.9\linewidth]{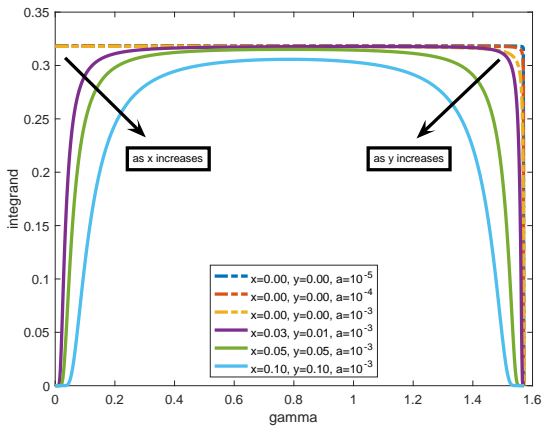}
\caption{Plot of the integrand of Eq.~\eqref{eq:ffun} versus $\gamma$
for different values of $x$, $y$, and $a$. }
\label{fig:integrand}
\end{figure}

At this stage, it is not strictly necessary to evaluate $f$ quickly because 
the time invested in computing $f$ does not affect the speed of the subsequent 
evaluation of $p$.
However, computational efficiency is always appreciated.
Therefore, we want to select an appropriate number of quadrature points
to speed up computations. For this goal, it is instructive to plot the integrand
of Eq.~\eqref{eq:ffun} for some chosen values of $x,y,$ and $a$.
In Fig.~\ref{fig:integrand} we observe that, as $x$ increases, the upper left 
corner smoothens.
Similarly, the upper right corner smoothens as $y$ increases. When $y$ is small,
the variable $a$ also affects the curvature of the upper right corner. 
For other values of $y$ the qualitative impact of $a$ is negligible (not shown).

\begin{figure}[htb!]
\centering
\includegraphics[width=0.9\linewidth]{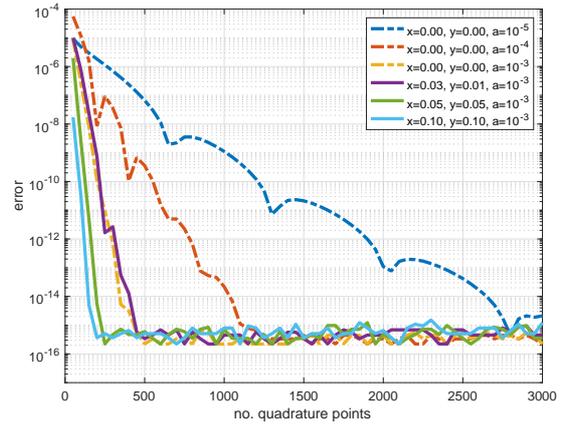}
	\caption{Gauss quadrature error in approximating {$f$}.
	The number of quadrature points necessary to achieve machine precision 
	depends on the values of $x$, $y$, and $a$.}
\label{fig:quaderror}
\end{figure}

From these observations we can speculate that approximating 
{$f$} becomes particularly challenging when $x$ or $y$ 
are close to zero.
This is confirmed in a numerical experiment, where we compare the difference 
between the value of $f$ approximated with \texttt{nQ} and \texttt{nQ+50} Gauss 
quadrature points (with \texttt{nQ} = 50, 10, \dots, 3000). 
The results are plotted in Fig.~\ref{fig:quaderror}.
The most challenging integral arises with $a=10^{-5}$ and $x = y = 0$. 
In this case, it takes more than 2500 Gauss quadrature points to approximate 
$f$ to machine precision. The number of
quadrature points necessary to achieve machine precision decreases drastically
if $x$, $y$, or $a$ increase.

After extensive numerical investigations, we decided to employ the following 
strategy to efficiently and accurately approximate $f$. 
If $x\geq 0.05$ and $y\geq 0.05$, we use 250 Gauss quadrature points. 
Otherwise, we use 700 Gauss quadrature points if $a\geq10^{-4}$ and
2500 Gauss quadrature points if $a<10^{-4}$.

\subsection{Results}
\label{sec:approxResults}
At this point, we can finally construct the trivariate polynomial $p$.
{In this example, we set the parameter \texttt{`eps'} to 
$10^{-11}$.}

\begin{figure}[htb!]
\centering
\includegraphics[width=0.99\linewidth]{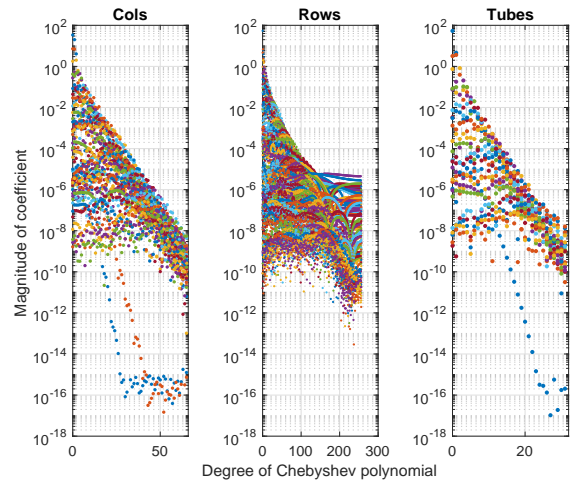}
\caption{Coefficients of the polynomials $c_i$ (Cols), $s_j$ (Rows), and $t_k$ 
	(Tubes) from Eq.~\eqref{eq:Chebfun3}}
\label{fig:plotcoeffs}
\end{figure}

After a roughly 140-minute-long computation on a standard laptop, mostly due 
to evaluating $f$ on interpolation points, Chebfun returns the interpolant $p$ 
given in Eq.~\eqref{eq:approximant}.
In reference to Eq.~\eqref{eq:Chebfun3}, its core $C_T$ is a 
$40 \times 74 \times 24$ tensor, and the $c_i$, $s_j$, and $t_k$
are univariate polynomials of degree 66, 257, and 32, respectively.
We can gather some extra information by plotting the Chebyshev coefficients 
of these polynomials (using the Chebfun command \texttt{plotcoeffs}). 
The result is displayed in Fig.~\ref{fig:plotcoeffs}. 
We observe that the coefficients of $c_i$ and $t_k$ decay exponentially,
whereas the coefficients of (some of the) $s_j$ reach a plateau. 
This indicates that the difficulty in approximating $f$ is mostly due to a 
nonanalytic behavior of $f$ in the $y$ direction.

\begin{figure}[htb!]
\centering
\includegraphics[width=0.9\linewidth]{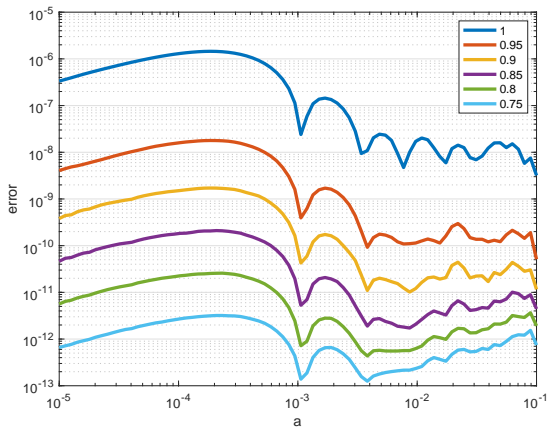}
\caption{Quantiles of the sampled error defined in 
	Eq.~\eqref{eq:samplederror}.}
\label{fig:abserrRe}
\end{figure}

To assess the accuracy of the constructed Chebfun3-based approximation, we 
consider the set of values of $b = -5, -5+1/20, \dots, -1$ and sample the error
\begin{equation}\label{eq:samplederror}
	\vert f(x,y,10^b) - \exp(p(x,y,b) \vert
\end{equation}
on the $(x,y)$-grid
\begin{multline}\label{eq:errorgrid}
	G=\{0,\frac{1}{40}, \frac{2}{40}, \dots, 3,3 +\frac{1}{20}, 
	3+\frac{2}{20} \dots, 6\} \\
	\times\{0,\frac{1}{40}, \frac{2}{40}, \dots, 3, 3+\frac{7}{60}, 
	3+\frac{14}{60}, \dots, 10\}\,.
\end{multline}
In Fig.~\ref{fig:abserrRe} we plot the statistics of these sampled errors.
The 1-quantile corresponds to the maximum error.
We observe that, in the worst case scenario, the Chebfun-based approximation
has 5 digits of accuracy, and that for $a>10^{-3}$ the number of exact digits 
is 6.
However, we also observe that the error is less 
than $10^{-8}$ for 95\% of the points in $G$ .

In certain physical situations it is advantageous to solve the RT equation in 
optical depth scale instead of geometric depth scale 
\citep[e.g.,][]{Janett17a,JanettPaganini18}. After
performing the required change of variables, the key physical quantity for
the solution of the RT equation is no longer the emission coefficient, but
rather the so-called source function (i.e., the ratio between the emission 
and absorption coefficients). %ALSO ASK THIS

\begin{figure}[htb!]
\centering
\includegraphics[width=0.9\linewidth]{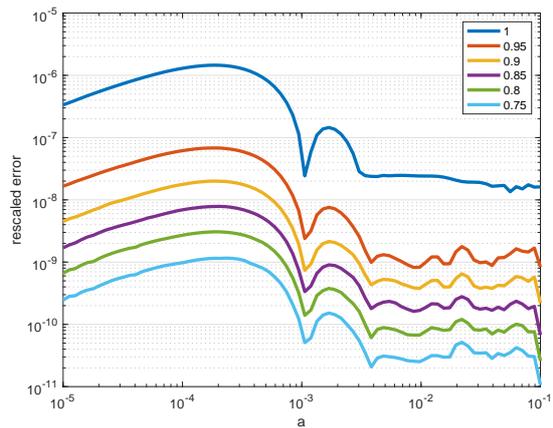}
\caption{Quantiles of the sampled rescaled error defined in 
	Eq.~\eqref{eq:samplederrorscaled}.}
\label{fig:abserrg}
\end{figure}

This calls for the calculation of the following {function
\begin{equation}
\label{eq:gfunc}
	g(x,y,a) = \frac{f(x,y,a)}{\phi(u)} \, ,
\end{equation}
}where the absorption profile $\phi$ is given by
\begin{equation}
	\phi(u) = \frac{1}{\sqrt{\pi}} H(a,u)\, .
\end{equation}
It is thus also interesting to investigate the accuracy of the proposed 
approximation for 
$g(x,y,a)$ 
For the sampling values of $x$, $y$, and $a$ used in this section, 
$\phi(u)\in [10^{-8}, 10^{-3}]$. Therefore, one could wonder whether 
the function $\phi(u)$ could nullify our efforts to approximate 
$\bigl[R_{\mathrm{II-AA}}\bigr]_{ij}(u^\prime,u)$ accurately. 
Luckily, this turns out not to be the case, as we can observe
in Fig.~\ref{fig:abserrg}, where we plot the values of 
\begin{equation}
	\label{eq:samplederrorscaled}
	\frac{\vert f(x,y,10^b) -  \exp(p(x,y,b) \vert}{\phi(u)}.
\end{equation}
We therefore conclude that the approximation presented here is extremely 
accurate even after dividing by $\phi(u)$.

\subsection{Evaluation of the polynomial approximation}
\label{sec:cheb3evaluation}

It is important to realize that, although it may take some time to construct
a Chebfun3 object, the subsequent evaluation of $p$ is very fast. 
To evaluate it at a point $(\tilde x, \tilde y, \tilde b)$ it is necessary to
first compute the values of the univariate polynomials $c_i$, $s_j$, and $t_k$
in $\tilde x, \tilde y,$ and $\tilde b$, respectively. This
can be done using 1D Clenshaw recurrence 
\citep{Clenshaw55, driscoll2014chebfun}, which is numerically stable and fast;
its time complexity is linear in terms of the degree of the polynomials.
Then, the values $c_i(\tilde x)$, $s_j(\tilde y)$, $t_k(\tilde b)$
are organized in vectors $c(\tilde x)$, $s(\tilde y)$, and $t(\tilde b)$.
Finally, the value $p(\tilde x, \tilde y, \tilde b)$
is obtained by computing the modal products
\cite[p. 727]{Golub13} of the core tensor $C_T$ with $c(\tilde x)$, 
$s(\tilde y)$, and $t(\tilde b)$. 
{These operations are also fast because, internally, they call 
high-performance implementations of Basic Linear Algebra Subroutines (BLAS) 
tuned by computer vendors for maximal speed and efficiency.}

To make an illustrative example, we 
{fixed} the value of $b$ and {evaluated} $p(x,y,b)$ at $10^5$
random points $(x,y)\in[0,5]\times[0,5]$. Such an experiment % 
{took} roughly {12} seconds.
The evaluation is dramatically faster if the points $(x,y)$ lie on a regular 
grid.
In this case, one can exploit the special structure of the regular grid, and 
evaluating $p$ on $10^6$ points takes only a fraction of a second. 
However, in the applications considered in this work, it is very rarely the 
case that one evaluates \eqref{eq:ffun} on a regular grid.
Therefore, the previous experiment with random points is more 
enlightening. 

{
\begin{remark}
{In practice, the emission coefficient must be evaluated through Eq.~\eqref{eq:emisPolar} at each individual 
spatial point. This requires the}
evaluation of $p(x,y,b)$ 
for {many} values of $x$ and $y$, but with $b$ fixed. % 
In this case, it is convenient to extract
a Chebfun2 object from $p$, that is, its equivalent bivariate counterpart
obtained by fixing the value of $b$. This Chebfun2 object can be
computed with the simple
Chebfun-command \verb!p2 = p(:,:,b)!. The resulting bivariate
polynomial $p_2$ returns exactly the same values of $p$, that is,
$p_2(x,y) = p(x,y,b)$, but it is much faster (because it does not
need to re-evaluate the polynomials $t_k$ in Eq.~\eqref{eq:Chebfun3}).
For instance, evaluating $p_2$ at $10^5$
random points $(x,y)\in[0,5]\times[0,5]$ takes only roughly 4 seconds.
\end{remark}
}

\section{{Fast and accurate approximation of $h$}}
\label{sec:FVoigt}
{
In this section we discuss how to approximate the function introduced in 
Eq.~\eqref{eq:hfun}, in close analogy to the previous section. 
We first point out that
the integrand contains the associated dispersion profile, which is an
odd function in its second argument.
For this reason, we can likewise} restrict our considerations to the positive
quadrant $x,y\geq0$.
Because the function $h$ also exhibits a super exponential decay in the 
variable $x$, we can restrict our considerations to the interval $x\in[0,6]$.
Finally, the main difference between \eqref{eq:hfun} and \eqref{eq:ffun}
is that the associated dispersion function vanishes when its second argument 
is zero, that is, $L(\cdot,0) = 0$. 
In light of these considerations, we compute two different approximations of 
\eqref{eq:hfun}: one for $y\in [10^{-13}, 10^{-1}]$ and one for 
$y\in [10^{-1},10]$.

For the regime $y\in [10^{-1},10]$ we employ the same strategy used in 
Sect.~\ref{sec:approximation}, and construct a trivariate polynomial $q$ 
such that
\begin{equation}
	\label{eq:im_appr_ylarge}
	q(x,y,b) \approx \log h(x,y,10^b)\,.
\end{equation}
On the other hand, for the regime $y\in [10^{-13}, 10^{-1}]$ it is more 
convenient to construct a trivariate polynomial $\tilde q$ such that
\begin{equation}
	\label{eq:im_appr_ysmall_nonlog}
	\tilde{q}(x,y,b) \approx h(x,y,10^b) \, .
\end{equation}
Finally, in the regime $y\in [0, 10^{-13}]$, the function $h$ satisfies 
$h < 4.1\cdot 10^{-13}$, and it can be simply approximated by the 
zero-constant function.

\begin{figure}[htb!]
\centering
\includegraphics[width=0.9\linewidth]{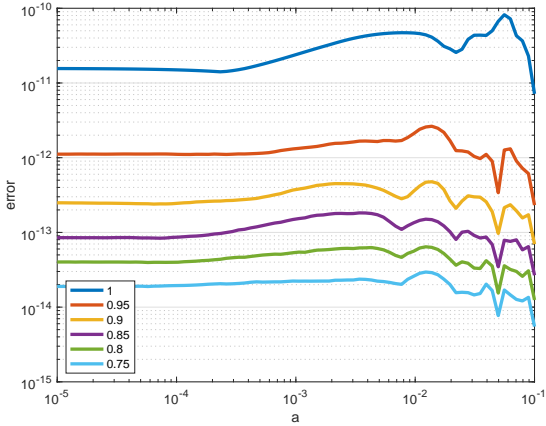}
\caption{Quantiles of the sampled error of Eq.~\eqref{eq:im_appr_ylarge}.}
\label{fig:abserrImYlarge}
\end{figure}

\begin{figure}[htb!]
\centering
\includegraphics[width=0.9\linewidth]{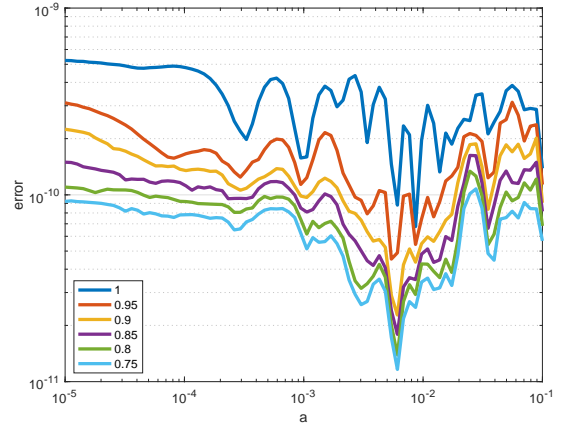}
\caption{Quantiles of the sampled error of Eq.~\eqref{eq:im_appr_ysmall_nonlog}.}
\label{fig:abserrImYsmall_nonlog}
\end{figure}

Similarly to Sect.~\ref{sec:approximation}, the approximants 
\eqref{eq:im_appr_ylarge} and \eqref{eq:im_appr_ysmall_nonlog}
can be computed using Chebfun.
The approximant $q$ is a trivariate polynomial in the Tucker form whose core 
tensor is of size $18 \times 33 \times 11$ with polynomials $c_i$, $s_j$, and 
$t_k$ of degree 28, 127, and 40, respectively.
Also, the approximant $\tilde q$ has trilinear rank $(6, 21, 17)$ with 
polynomials $c_i$, $s_j$, and $t_k$ of degree 30, 257, and 41, respectively.
In Figs.~\ref{fig:abserrImYlarge} and \ref{fig:abserrImYsmall_nonlog} we
display the quantiles of the error described in Sect.~\ref{sec:approxResults}
(without {dividing by $\phi(u)$}).
These figures show that the constructed approximation is extremely accurate.

\section{Comparison on a physical application}
\label{sec:application}

To make a comparison of practical interest, we 
consider the evaluation of the 
integral over $u^\prime$ contained in Eq.~\eqref{Eq:emis}. 
{This corresponds to the unpolarized case, 
in which only the function $f$ appears in the problem. 
The extension to the polarized case, in which both 
functions $f$ and $h$ are involved (see Sect.~\ref{sec:intro} and Appendix~\ref{sec:Eval_append}), is discussed at the end of this section. } 

{In this experiment, 
we assume  $\mathcal{P}(\vec{\Omega}^\prime,\vec{\Omega}) = 1$ 
{in Eq.~\eqref{Eq:redisfac}} (isotropic scattering),
 and} we introduce the mean intensity
\begin{equation}
\label{eq:meanint}
	J(u) = \frac{1}{4\pi} \oint {\rm d} \vec{\Omega} \, I(u,\vec{\Omega}) \, .
\end{equation} 
For simplicity, we consider a single spatial grid point in the atmospheric 
model.
Let $\{ u_i \}$, $i=1, \dots, {N_F}$, be the grid of reduced frequencies at 
that point.\footnote{We recall that, for a given frequency grid, the values
of the reduced frequencies $\{ u_i \}$ depend on the spatial point.} 
Our goal is to calculate, at all the emitted reduced frequencies $\{ u_i \}$, 
the integral (often referred to as scattering integral) 
{
\begin{equation}
\label{eq:scatint}
	{\mathcal I}(u_i) = \int \, {\rm d} u^\prime \,
	f\Bigl(\frac{u^\prime - u_i}{2}, \frac{u^\prime + u_i}{2}, a \Bigr) \, 
	J(u^\prime) \, , 
\end{equation}
}
which is related to the emission coefficient simply as
{
\begin{equation}
\label{eq:emiscat}
	\varepsilon_I(u_i,\vec{\Omega}) = k_L \, 
	\frac{\Gamma_R}{\Gamma_R + \Gamma_I + \Gamma_E} \,  {\mathcal I}(u_i) \, . 
\end{equation}
Since {$f$}} presents sharp variations with 
$u^\prime$, approximating integral \eqref{eq:scatint} with the trapezium rule 
on the grid $\{ u_i \}$ does not return accurate results.
A successful approach, in this case, is to construct an interpolant of $J$ 
and then perform the integration on a much finer grid.

Following this approach, let $\bigr\{b_{j}(u')\bigl\}$ be an interpolation 
basis such that $b_j(u_i) = \delta_{i j}$. 
Then, the interpolant of $J$ can be written as
\begin{equation}
	J(u^\prime) \approx \sum_{j=1}^{N_F} \mu_j \, b_j(u^\prime) \, ,
\label{Eq:rad_interp}
\end{equation}
where $\mu_j$ denote the interpolant coefficients ($\mu_j = J(u_j)$).
After this substitution, the scattering integral \eqref{eq:scatint} can be 
approximated by
{
\begin{equation}
	\mathcal{I}(u_i) \approx \sum_{j=1}^{N_F} 
	\mathrm{F}_{j i} \, \mu_j , \; (i=1,...,N_F), 
\label{Eq:freq_int_interp} 
\end{equation}
}where the quadrature weights are given by {
\begin{equation}
	 \mathrm{F}_{j i} = \int {\rm d} u^\prime \,  
      f\Bigl(\frac{u^\prime - u_i}{2}, \frac{u^\prime + u_i}{2},a\Bigr) \, b_j(u^\prime) \, .
\label{Eq:scat_quad_weights}
\end{equation} 
}
From Eq.~\eqref{Eq:freq_int_interp} we conclude that the main computational 
cost is approximating the quadrature weights {$\mathrm{F}_{ji}$}.
Indeed, once these have been found, integral \eqref{eq:scatint}
can be computed almost instantaneously with formula \eqref{Eq:freq_int_interp}.

Of course, the quadrature weights {$\mathrm{F}_{j i}$} depend on the basis 
functions $\bigr\{b_{j}(u^\prime)\bigl\}$ used in Eq.~\eqref{Eq:rad_interp}. 
In the past, different authors have recommended using cardinal natural 
interpolatory cubic splines 
\citep[e.g.,][]{Adams+71,Gouttebroze86,Uitenbroek89}. 
The order of convergence of cubic splines is quartic, but 
their interpolatory basis functions are oscillatory and have global support.
This introduces an extra difficulty in evaluating the weights {$\mathrm{F}_{ji}$}.
Developing an efficient algorithm to compute these weights is
beyond the scope of this work, and we postpone it to future research.
In this work, we consider linear B-splines instead.
Linear B-splines are piecewise linear functions and have compact support, 
which simplifies the task of computing {$\mathrm{F}_{ji}$}.

Assuming that the values $\{u_i\}$ increase monotonically, the linear B-spline 
associated with an internal frequency point
$u_j\in (u_1, u_{N_F})$ is
\begin{equation*}
	b_j(x) = 
	\begin{cases}
		\frac{x-u_{j-1}}{u_j-u_{j-1}} \, , & 
		x\in (u_{j-1},u_j] \\
		\frac{u_{j+1}-x}{u_{j+1}-u_{j}} \, , & 
		x\in (u_{j},u_{j+1}] \\
		0\,, & \text{otherwise.}
	\end{cases}
\end{equation*}
There is some freedom in defining the linear B-splines associated with the
first and the last frequency points. 
Given that in the applications considered below we will only consider basis 
functions associated with internal points, this choice is not particularly
relevant for the scope of this work.
However, one can, for instance, consider extension by a constant value,
and use
\begin{equation*}
	b_1(x) = 
	\begin{cases}
		1 \, , & u \leq u_1 \\
		\frac{u_{2}-x}{u_{2}-u_{1}} \, , & x\in (u_{1}, u_{2}] \\
		0\,, & \text{otherwise}
	\end{cases}
\end{equation*}
and
\begin{equation*}
	b_{N_F}(x) = 
	\begin{cases}
		\frac{x-u_{N_F-1}}{u_{N_F}-u_{N_F-1}} \, , & x\in (u_{N_F-1},u_{N_F}]\\
		1 \, , & x> u_{N_F} \\
		0\,, & \text{otherwise,}
	\end{cases}
\end{equation*}
respectively.

In {principle}, the interval of integral 
\eqref{Eq:scat_quad_weights} is the whole real line. 
However, since $b_j$ has compact support, we can restrict the integration 
interval to $(u_{j-1},u_{j+1})$. 
Moreover, the {function $f$} decays
super-exponentially as the quantity $\vert u^\prime - u_i \vert$ 
increases (see Fig.~\ref{fig:realRII}).
Therefore, it is necessary to integrate \eqref{Eq:scat_quad_weights} only
for $u^\prime$ in
\begin{equation}
(u_{j-1},u_{j+1})\cap (u_i - 12, u_i + 12)\,.
\label{eq:interval}
\end{equation}
An efficient way to compute integral~\eqref{Eq:scat_quad_weights} is to further 
split {interval}~\eqref{eq:interval} at $u_i$ and $u_j$ and use 
different Gauss quadrature rules in each subinterval because $b_j$
{and }
{$f$} 
are not smooth functions. 
Without going into details, {for each non-empty interval given by Eq.~\eqref{eq:interval}}, this approach is expected to require at least 
(roughly) 20 quadrature points in total. 

In this numerical experiment, we evaluate {$f$} on these quadrature points using Chebfun-based 
approximations.
The goal is to make a comparison in terms of speed and accuracy with
direct computations of Eq.~\eqref{eq:ffun} based on quadrature rules. 

To give this experiment a realistic application flavor, we use a grid of 
reduced frequencies considered in a realistic RT problem. 
In particular, our grid is the one used in the RT investigation of the 
Mg~{\sc ii} $k$ line presented in \cite{Alsina+16}, specifically corresponding 
to the height point at 2075\,km in the atmospheric model C of 
\cite{Fontenla+93}.
This reduced frequency grid consists of 109 points spanning the range between
$u = -333.5$ and $u = 328$. 
It contains 23 equispaced points in the core of the line, with the remaining 
points being logarithmically spaced outside this range. 

For this grid, we consider every nonempty interval of the form
\eqref{eq:interval} (which in this case are 4005), and for each of these we
collect 10, 20, {and 40 quadrature points (for a total of 40050, 80100, 
and 160200)}.  
 Using 10 points corresponds to a coarse approximation of 
 \eqref{Eq:scat_quad_weights}, {whereas 
between 
20 and 40 Gauss quadrature points should be considered for
higher accuracies.}  
To simplify the numerical experiment, we replace composite Gauss quadratures 
with the trapezium rule. 

For each of these four cases, we measure the time necessary to evaluate 
\eqref{eq:ffun} at all quadrature points using Chebfun-based approximations 
with parameter
{\texttt{eps} = $10^{-5},10^{-6},10^{-7}, 10^{-8}, 10^{-9}, 10^{-10}, 10^{-11}$,}
and using a direct Gauss-quadrature
approximation\footnote{This integration is performed with a highly efficient 
and fully vectorized code in \textsc{Matlab}.} with
{8, 16, 32, 64, 128, 256, 512}
points.  
{For Chebfun-based solutions, we consider both
approximations obtained by evaluating the Chebfun3 {object}
\eqref{eq:Chebfun3} as well as approximations obtained by
evaluating a Chebfun2 {object} generated by fixing $b = \log_{10}{a}$
in the original Chebfun3 {object}. These two Chebfun-based approximations
return exactly the same values, but have different execution times,
because the original Chebfun3 {object} inevitably re-evaluates the polynomials
$t_k$ (see Eq.~\eqref{eq:Chebfun3}).}
We also compute the maximum error for each of these approximations using 
self- and cross-comparison.
Self-comparison means that we consider convergence of Chebfun-solutions to the 
values obtained using the Chebfun-solution with \texttt{eps} = $10^{-11}$,
and the convergence of quadrature-solutions to the quadrature-solution with 
512 Gauss quadrature points.
Cross-comparison means that we consider convergence of Chebfun-solutions to 
the quadrature-solution with 512 Gauss quadrature points, and convergence of
quadrature-solutions to the Chebfun-solution with \texttt{eps} = $10^{-11}$.

\begin{figure}[htb!]
\centering
\includegraphics[width=0.9\linewidth]{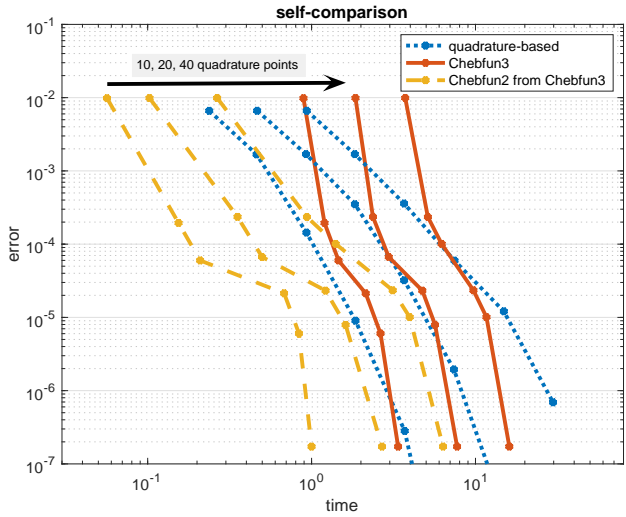}
\caption{Self-comparison error versus computational time of Chebfun-based and 
	quadrature-based approximations of % 
	{$f$} on $4005\times 10, 20, 40$ quadrature points.
	{The circles on the curves for quadrature-solutions indicate, 
	for decreasing error, the results obtained taking 8, 16, 32, 64, 128, and 256 
	quadrature points. The circles on the curves for Chebfun solutions indicate, 
	for decreasing error, the results obtained taking 
	\texttt{eps} = $10^{-5},10^{-6},10^{-7}, 10^{-8}, 10^{-9}$, and $10^{-10}$.}}
\label{fig:self-comparison}
\end{figure}

\begin{figure}[htb!]
\centering
\includegraphics[width=0.9\linewidth]{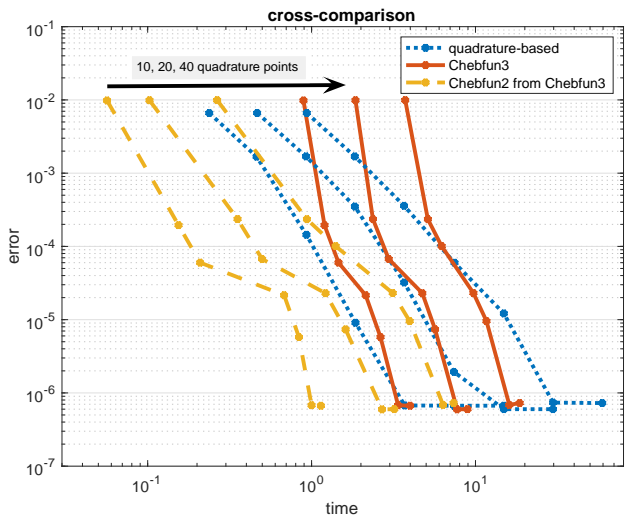}
\caption{Cross-comparison error versus computational time of Chebfun-based and 
	quadrature-based approximations of % 
	{$f$} on $4005\times 10, 20, 40$ quadrature points.
	{The circles on the curves for quadrature-solutions indicate, 
	for decreasing error, the results obtained taking 8, 16, 32, 64, 128, 256, and 512  
	quadrature points. The circles on the curves for Chebfun solutions indicate, 
	for decreasing error, the results obtained taking 
	\texttt{eps} = $10^{-5},10^{-6},10^{-7}, 10^{-8}, 10^{-9}, 10^{-10}$, and $10^{-11}$.}
	}
\label{fig:cross-comparison}
\end{figure}

In Figs.~\ref{fig:self-comparison} and \ref{fig:cross-comparison}, we display 
the error versus computational time for Chebfun-based and quadrature-based 
solutions.
{In these simulations, the error ranges within the interval
{$[10^{-6},10^{-2}]$}, which is the most relevant for practical applications. 
We observe that {the} Chebfun2-based solutions
are notably faster than quadrature-based solutions. For instance,
{the} slowest (and most accurate) {Chebfun2-based object} requires essentially
the same computational time {as} the quadrature-based solution obtained with 30
Gauss quadrature points (and it is 3 orders of magnitude more accurate).
We also note that the Chebfun3-{based approximations} 
are as fast as quadrature-based solutions for errors smaller than $2\cdot 10^{-4}$,
if not faster. 
Finally, we observe that, in the cross-comparison plot,
Chebfun- and quadrature-based solutions converge to each other until the error is  
{below $10^{-6}$, }
at which point the {``convergence lines''} plateau. 
This is expected because Chebfun solutions can guarantee a precision of 
roughly $10^{-7}$ (cf. Fig.~\ref{fig:abserrRe}).} 

{\begin{remark}
In this numerical experiment, we have % 
{analyzed}
the time required
to compute the emission coefficient % 
{at}
a single spatial point and
in the absence of polarization. 
{We now comment on these two simplifications.}

As mentioned previously, standard 1D 
semi-empirical models of the solar atmosphere easily contain 100 spatial points 
or more. Since the values of the reduced frequencies {and of the damping parameter 
vary with the spatial point},
a simulation with 100 spatial points requires computing 100 different sets of 
quadrature weights $\{F_{ji}\}$ \eqref{Eq:scat_quad_weights}, and thus 
the gain in computational time is multiplied by 100 if these are not computed in parallel.

The computation of the emission coefficient {for polarized radiation}
is fully analogous to the unpolarized case.
Deferring the details to Appendix~\ref{sec:Eval_append}, 
we mention that the polarized case {additionally} requires the computation of quadrature weights
of the form \eqref{Eq:scat_quad_weights} % 
{replacing the}
function $f$ by the function $h$. 
{Repeating the previous experiments 
for function $h$
 would simply produce plots similar to 
Figs.~\ref{fig:self-comparison} and \ref{fig:cross-comparison}
because the approximation of $h$
is completely analogous to the approximation of $f$.} 
On top of that, {in the presence of a magnetic field---or 
when considering an atomic model more complex than one with two levels---
multiple sets of functions $f$ and $h$ with different shifts in their arguments
are involved and a separate quadrature weight must be determined for each of them
(see Eqs.~\eqref{eq:fweight} and \eqref{eq:hweight}).  
Clearly, the overall computation time scales with the number of quadrature 
weights that are required.}  
\end{remark}
}

\section{Conclusion}
\label{sec:concl} 
{We have presented a new method for
approximating the functions $f$ and $h$ 
that appear in the angle-averaged  
redistribution 
{matrix for polarized radiation} 
$\bigl[\mathrm{R}_{\mathrm{II-AA}}\bigr]_{i j}$.} 
The new method uses low-rank approximation and Chebyshev polynomials to
construct functions that approximate $f$ and $h$. These approximating
functions can be evaluated quickly using Clenshaw recurrence. The level 
of accuracy does not deteriorate when dividing these
approximating functions by the absorption profile to compute the source
function. Numerical experiments performed in a realistic scenario show that
our approach permits {significantly} faster computations than
standard algorithms based on quadrature rules{, while
achieving similar or higher accuracies}. 
{In addition, at any given height point (where the parameter $a$
is fixed), the evaluation can be carried out even faster using Chebfun2
objects generated as cross-sections of the main Chebfun3 object.}

{The} decrease in computational cost 
provided by this new {method} will be very useful for RT 
{investigations} of scattering polarization and the
Hanle and Zeeman effects in strong resonance lines, accounting for 
the impact of PRD phenomena, 
{either in physical situations for which an angle-dependent 
treatment is not presently feasible, or when rapid calculations at a low 
computational cost are required.} 
This method is expected to be particularly valuable when considering 
complex atomic models such as multi-term atoms or atoms with hyperfine
structure.

\appendix
\section{Expression of the $[R_\mathrm{II-AA}]_{ij}$ 
redistribution matrix for a two-level atom}
\label{app:A}
{As shown in detail in \citet{Alsina+17}, in the polarized case, the 
angle-averaged $[R_\mathrm{II-AA} ]_{ij}$ redistribution matrix for a 
two-level atom with an unpolarized and infinitely sharp lower level, in the 
presence of magnetic fields, is given by}
\begin{align}
	\label{Eq:redismat}
	& \left[ R_{\mathrm{II-AA}}(u^\prime,\vec{\Omega}^\prime,u,\vec{\Omega}) 
	\right]_{i j} = \nonumber \\
	& \quad \sum_{K K^\prime Q}
	\bigl[\mathcal{R}_\mathrm{II-AA} \bigr]^{K K^\prime}_Q \! (u^\prime,u) \, 
	\mathcal{P}^{K K^\prime}_Q(\vec{\Omega}^\prime,\vec{\Omega})_{ij} \, ,
\end{align}
where the indices $K$ and $K'$ can take values 0, 1, and 2, while 
$Q$ can take integer values between $-K_\mathrm{min}$ and $+K_\mathrm{min}$, 
with $K_\mathrm{min} = \mathrm{min}(K,K')$.

The $4\times4$ matrices $\mathcal{P}^{KK'}_Q(\vec{\Omega}',\vec{\Omega})_{ij}$
generalize to the polarized case (within the framework of the irreducible 
spherical tensors formalism) the angular phase function appearing in 
Eq.~(\ref{Eq:redisfac}).
In a reference system such that the $z$-axis (quantization axis) is directed along 
the magnetic field, they are given by
\begin{align}
	\label{Eq:GeomRedisMat}
	\mathcal{P}^{K K^\prime}_Q(\vec{\Omega}^\prime,\vec{\Omega})_{ij} = 
	(-1)^Q \mathcal{T}^K_Q(i,\vec{\Omega}) 
	\mathcal{T}^{K^\prime}_{-Q}(j,\vec{\Omega}^\prime) \, ,
\end{align}
where $\mathcal{T}^K_Q$ is the so-called polarization tensor 
\citep[see][]{BLandiLandolfi04}.
The expression of $\mathcal{P}^{K K^\prime}_Q$ in an arbitrary reference system 
can be found through simple rotations \citep[e.g.,][]{Alsina+17}.

The quantities $\left[{\mathcal R}_{\mathrm{II-AA}} \right]^{K K^\prime}_Q$
are given by \citep[see][]{Alsina+17} %{
\begin{align}
	& \left[ {\mathcal R}_{\mathrm{II-AA}} \right]^{K K^\prime}_Q \! 
	(u^\prime,u) \, = \nonumber \\
	& \sum_{\substack{M_u M_u^\prime M_\ell M_\ell^\prime \\ 
	p \, p^\prime p^{\prime \prime} p^{\prime \prime \prime}}}
	\frac{\Gamma_R}{\Gamma_R + \Gamma_E + \Gamma_I + \mathrm{i}\, \omega_L 
	g_{u} Q} \nonumber \\
	& \; \, \times {\mathcal C}_{K K^\prime Q M_u M_u^\prime M_\ell 
	M_\ell^\prime\, p p^\prime p^{\prime \prime} p^{\prime \prime \prime}} 
	\notag \\
	& \; \, \times \frac{1}{2\pi} \int_0^\pi \!\!\mathrm{d}\Theta
	\exp\Biggl[-\biggl(\frac{u^\prime - u - \Delta_{M_\ell M^\prime_{\ell}}}{2\sin(\Theta/2)} \biggr)^2 \Biggr]
	\notag \\
	& \, \times \frac{1}{2} \Biggl\{ 
	W\biggl(\frac{a}{\cos(\Theta/2)}, \frac{u + u^\prime + \Delta_{M^\prime_{u} M_\ell} + \Delta_{M^\prime_{u} M^\prime_{\ell}}}{2 \cos(\Theta/2)} \biggr) \notag \\
	 & \; \; \; +
	W\biggl(\frac{a}{\cos(\Theta/2)}, \frac{u + u^\prime + \Delta_{M_{u} M_\ell} + \Delta_{M_{u} M^\prime_{\ell}}}{2 \cos(\Theta/2)} \biggr)^\ast 
 \Biggr\} \, ,
 \label{eq:R_Decomp}
\end{align}
where we have introduced the Faddeeva function
\begin{equation*}
W(a,x) = H(a,x) + \mathrm{i} \, L(a,x) \, , 
\end{equation*}
%}
and $M_u$, $M_u^\prime$, $M_\ell$, $M_\ell^\prime$ are the magnetic quantum 
numbers corresponding to the various substates of the upper (subscript $u$) 
and lower (subscript $\ell$) levels.
The quantity $g_u$ is the Land\'e factor of the upper level, $\omega_L$ is the 
angular Larmor frequency (which depends on the magnetic field intensity), and 
$\mathcal{C}$ is a factor related to the coupling of the various magnetic
quantum numbers \citep[for its explicit expression see][]{Bommier97b}.
The integers $p, p^\prime, p^{\prime \prime}$, and $p^{\prime \prime \prime}$
it contains range from $-1$ to $1$. 
{The % 
frequency shifts of the Zeeman transition between the upper level with $M_u$ and
the lower level with $M_\ell$, with respect to the line-center frequency $\nu_0$
is given by
}{
\begin{equation}
	\Delta_{M_u M_\ell} = \frac{E(M_u)/h - E(M_\ell)/h - \nu_0}{\Delta\nu_D} \, , 
	\label{Eq:mag_shift_1} 
\end{equation}
whereas the frequency splitting between two lower levels with $M_\ell$ and $M_\ell^\prime$
is given by
\begin{equation}
	\Delta_{M_\ell M^\prime_{\ell}}  = \frac{E(M_\ell)/h - E(M_\ell^\prime)/h}
	{\Delta\nu_D} \, ,
	\label{Eq:mag_shift_2}
\end{equation}
where $E(M)$ is the energy of a given magnetic sublevel, $h$ is the Planck 
constant, and $\Delta\nu_D$ is the 
Doppler width of the line.}

{One can immediately realize that the quantity shown in Eq.~\eqref{eq:R_Decomp} may 
be expressed in terms of the functions $f$ and $h$ given in Eqs.~\eqref{eq:ffun} and \eqref{eq:hfun}
as} { 
\begin{align}
	& \left[ {\mathcal R}_{\mathrm{II-AA}} \right]^{K K^\prime}_Q \! 
	(u^\prime,u) \, = \nonumber \\
	& \sum_{\substack{M_u M_u^\prime M_\ell M_\ell^\prime \\ 
	p \, p^\prime p^{\prime \prime} p^{\prime \prime \prime}}}
	\frac{\Gamma_R}{\Gamma_R + \Gamma_E + \Gamma_I + \mathrm{i}\, \omega_L 
	g_{u} Q} \nonumber \\
	& \; \, \times {\mathcal C}_{K K^\prime Q M_u M_u^\prime M_\ell 
	M_\ell^\prime\, p p^\prime p^{\prime \prime} p^{\prime \prime \prime}} 
	\notag \\
	& \; \, \times \frac{1}{2} 
	\biggl[ 
	f(x_{\ell \ell^\prime},y_{u^\prime \ell \ell^\prime},a) + 
	f(x_{\ell \ell^\prime},y_{u \ell \ell^\prime}, a) \nonumber \\
	& \qquad \quad + \mathrm{i} 
	\Big( h(x_{\ell \ell^\prime},y_{u^\prime \ell \ell^\prime},a)  -  
	h(x_{\ell \ell^\prime},y_{u \ell \ell^\prime},a) \Big) 
	\biggr] \, ,
 \label{eq:R_Decomp_functions}
\end{align}
{where the dependence on the involved states is included in the variables}
\begin{align}
	& x_{\ell \ell^\prime} = \frac{u^\prime - u - 
	\Delta_{M_\ell M_{\ell^\prime}}}{2} \, , \label{eq:xexpl}\\
	& y_{u \ell \ell^\prime} = \frac{u^\prime + u + \Delta_{M_u M_\ell} + 
	\Delta_{M_u M_{\ell^\prime}}}{2} \, . \label{eq:yexpl} 
\end{align}
}
{It is interesting to observe that if the 
magnetic fields and the polarization of the radiation are neglected, the expressions of Eqs.~(\ref{Eq:redisfac})
and (\ref{Eq:RII-AA_unpol}) are recovered. 
When there is no magnetic splitting of the Zeeman sublevels, 
the arguments of the various functions $f$ and $h$ are all equal for any 
given $u$ and $u^\prime$. 
Thus, the various contributions from the functions $h$ cancel each other, and 
the imaginary part of Eq.~(\ref{eq:R_Decomp}) vanishes.} 

{Moreover, when such splittings are absent, it is possible to perform 
the following sum over the magnetic quantum numbers,}  
\begin{align}
	& \sum_{\substack{M_u M_u^\prime M_\ell M_\ell^\prime \\ 
	p \,p^\prime p^{\prime \prime} p^{\prime \prime \prime}}}
	\mathcal{C}_{KK'Q M_u M_u^\prime M_\ell M_\ell^\prime p p' p'' p'''} = 
	\nonumber \\
	& \qquad \qquad \qquad = \delta_{KK'} W_K(J_\ell, J_u) \, ,
\end{align}
where $W_K$ is %the so-called polarization factor 
a factor characterizing the polarizability of the considered line 
\citep[see][]{BLandiLandolfi04}, and $J_u$ and $J_\ell$ are the total angular 
momenta of the upper and lower level, respectively.
Observing that in the absence of magnetic fields $\omega_L = 0$, the quantity 
$\left[\mathcal{R}_{\mathrm{II-AA}}\right]^{K K'}_Q$ thus reduces to
\begin{align}
\label{eq:RedZeroField}
	& \left[\mathcal{R}_{\mathrm{II-AA}}\right]^{K}(u^\prime,u) =   
	\frac{\Gamma_R}{\Gamma_R + \Gamma_E + \Gamma_I} \nonumber \\
	& \qquad \times W_K(J_\ell,J_u) \, f(x,y,a) \, .
\end{align}

In the particular case of a transition with $J_\ell = 0$ and $J_u =~1$
(whose results correspond to the semi-classical picture), the factors 
$W_K$ are all equal to unity, so that \eqref{eq:RedZeroField} has no dependence 
on $K$ and reduces to \eqref{Eq:redisfac}, valid in the unpolarized case. 
Summing over the components of the angular phase matrix and neglecting the 
polarization of the radiation field, one recovers the dipole scattering
angular phase function
\begin{align}
	& \sum_{KQ} \mathcal{P}^{KK}_Q(\vec{\Omega},\vec{\Omega}')_{00} = \\
	& \sum_{KQ} (-1)^Q \, \mathcal{T}^K_Q(0,\vec{\Omega}) \, 
	\mathcal{T}^K_{-Q}(0,\vec{\Omega}') = 
	\frac{3}{4} \left(1 + \cos^2{\Theta} \right) \, , \nonumber
\end{align}
with $\Theta$ the scattering angle. 
If, in addition, the incident radiation field is isotropic, then 
\eqref{eq:radtens} introduces a $\delta_{K 0} \, \delta_{Q 0}$ (i.e., the only 
nonzero multipolar component of the radiation field tensor is $J^0_0$). 
Thus, the only term of the previous sum that contributes to the emissivity is  
\begin{equation}
	\mathcal{P}^{00}_0(\vec{\Omega},\vec{\Omega}')_{00} = 1 \, ,
\end{equation}
which represents the angular phase function for isotropic scattering. 

\section{{Numerical evaluation of the emission 
coefficient in the polarized case}}
\label{sec:Eval_append}
The method introduced in Sect.~\ref{sec:application} for the numerical 
evaluation of the frequency integral 
{of} Eq.~\eqref{Eq:emis} can be easily generalized 
from the unpolarized to the polarized case  
({see} Eq.~\eqref{eq:emisPolar}). 
{As a first step, we substitute} Eq.~\eqref{Eq:GeomRedisMat} 
into Eq.~\eqref{Eq:redismat} and we introduce the radiation field tensor
\begin{equation}
	J^K_Q(u) = \frac{1}{4 \pi} \, \oint {\rm d}\vec{\Omega} \, \sum_{i=0}^3 
	\mathcal{T}^K_Q(i,\vec{\Omega}) \, I_i(u,\vec{\Omega}) \, . 
	\label{eq:radtens}
\end{equation}
We note that $J^0_0$ corresponds to the mean intensity given in 
Eq.~\eqref{eq:meanint}. 
The emission coefficient for the polarized case{, given in Eq.~\eqref{eq:emisPolar},} can 
{thus} be written as 
\begin{align}
	\varepsilon_i(u,\vec{\Omega}) = k_L \sum_{K Q} 
	\mathcal{T}^K_Q(i,\vec{\Omega})
	\sum_{K^\prime} \mathcal{I}^{K K^\prime}_Q(u) \, ,
\label{eq:emiscatPol}
\end{align}
where the $\mathcal{I}^{K K^\prime}_Q(u)$ represents an extension to the 
polarized case of the scattering integral of Eq.~\eqref{eq:scatint}:
\begin{align}
	\mathcal{I}^{K K^\prime}_Q(u) = \, & (-1)^Q \, \int_{-\infty}^\infty {\rm d}u^\prime \, 
	\left[\mathcal{R}_{\mathrm{II-AA}} \right]^{K K^\prime}_Q \! (u^\prime,u) 
	J^{K^\prime}_{-Q}(u^\prime) .
\label{eq:scatintPol}
\end{align}
{From Eq.~\eqref{eq:R_Decomp_functions} it is immediate to realize that 
quantities {of the form \eqref{eq:scatintPol}} {can} be expressed as linear combinations of functions of the 
form} 
\begin{align}
	\mathcal{F}^K_Q(u; \alpha) = (-1)^Q \int_{-\infty}^\infty\!\!\mathrm{d}u^\prime \,
	f(x_{\alpha}, y_{\alpha}, a) J^K_{-Q}(u^\prime) \,,
	\label{eq:fint} \\
	\mathcal{H}^K_Q(u; \alpha) = (-1)^Q \int_{-\infty}^\infty\!\!\mathrm{d}u^\prime \, 
	h(x_{\alpha}, y_{\alpha}, a) J^K_{-Q}(u^\prime) \,.
	\label{eq:hint}
\end{align}
{In order to simplify the notation, in the previous expressions
we have introduced the label $\alpha$ %, %{which refers} to the %scattering ``channel'' involving
%particular scattering processes involving %ASK LUCA
%term involving 
to indicate the set of 
states with magnetic quantum numbers $M_\ell$ (initial), $M_\ell^\prime$ (final), and $M_u$. The arguments in
the function $f$ and $h$ are $x_\alpha = x_{\ell \ell^\prime}$ and $y_\alpha = y_{u \ell \ell^\prime}$. } 
 
{In full analogy with} Sect.~\ref{sec:application}, 
given a reduced frequency grid $\{ u_i \}$ and an {interpolatory} basis 
{$\{b_j(u^\prime)\}$}, the interpolant for the $J^K_Q$ 
components can be written as
\begin{equation}
\label{eq:rad_tens_interp_multi}
	J^K_Q(u^\prime) \approx \sum_{j=1}^{N_F} \left[\mu^K_Q \right]_j \, 
	b_{j}(u') \, .
 \end{equation} 
{By using this interpolation, the quantities appearing in Eqs.~\eqref{eq:fint} and 
\eqref{eq:hint} can be approximated as}
\begin{align}
	\mathcal{F}^K_Q(u_i; \alpha) \approx (-1)^Q \sum_{j=1}^{N_F} 
	\left[\mathrm{F}\right]^{\alpha}_{j i} 
	\left[\mu^K_{-Q} \right]_j \label{eq:fdecomp} \, , \\
	\mathcal{H}^K_Q(u_i; \alpha) \approx (-1)^Q \sum_{j=1}^{N_F}
	\left[\mathrm{H}\right]^{\alpha}_{j i} 
	\left[\mu^K_{-Q} \right]_j \label{eq:hdecomp} \, ,
\end{align}
where we have introduced the weights 
\begin{align}
	\left[\mathrm{F}\right]^{\alpha}_{j i} = 
	\int_{-\infty}^\infty\!\! \mathrm{d}u^\prime f(x_{\alpha, i}, y_{\alpha, i}, a) \, b_j(u^\prime) \, , 
	\label{eq:fweight} \\
	\left[\mathrm{H}\right]^{\alpha}_{j i} = 
	\int_{-\infty}^\infty\!\! \mathrm{d}u^\prime h(x_{\alpha, i}, y_{\alpha, i}, a) \, b_j(u^\prime) \, , 
	\label{eq:hweight}   
\end{align} 
{in which} $x_{\alpha, i}$ and $y_{\alpha, i}$ are 
{variables $x_\alpha$ and $y_\alpha$}
{with $u = u_i$ in Eqs.~\eqref{eq:xexpl}
and \eqref{eq:yexpl}}. 
{As in Sect.~\ref{sec:application}, it is apparent that the evaluation of these weights represents the majority of the 
computational cost involved in the 
frequency integral for the polarized emission coefficient. 
This involves the computation of distinct $F$ and $H$ weights for each set of states
$\alpha$, which differ from each other in the evaluation of the functions $f$ and $h$ at different $x_{\alpha, i}$ and $y_{\alpha, i}$ points. }

\begin{acknowledgements}
{The work of B.H. was in part supported by a grant from IPM (No.~$99650033$). E.A.B and L.B. gratefully acknowledge financial support from
the Swiss National Science Foundation (SNSF) through grants No.~$200021\_175997$ and
CRSII$5\_180238$.} 
\end{acknowledgements}

% \bibliography{refs.bib}

\begin{thebibliography}{28}
\expandafter\ifx\csname natexlab\endcsname\relax\def\natexlab#1{#1}\fi

\bibitem[{{Adams} {et~al.}(1971){Adams}, {Hummer}, \& {Rybicki}}]{Adams+71}
{Adams}, T.~F., {Hummer}, D.~G., \& {Rybicki}, G.~B. 1971, J. Quant. Spectrosc.
  Radiat. Transfer, 11, 1365

\bibitem[{{Alsina Ballester} {et~al.}(2016){Alsina Ballester}, {Belluzzi}, \&
  {Trujillo Bueno}}]{Alsina+16}
{Alsina Ballester}, E., {Belluzzi}, L., \& {Trujillo Bueno}, J. 2016, ApJ, 831,
  L15

\bibitem[{{Alsina Ballester} {et~al.}(2017){Alsina Ballester}, {Belluzzi}, \&
  {Trujillo Bueno}}]{Alsina+17}
---. 2017, ApJ, 836, 6

\bibitem[{{Alsina Ballester} {et~al.}(2018){Alsina Ballester}, {Belluzzi}, \&
  {Trujillo Bueno}}]{AlsinaBallester+18}
---. 2018, \apj, 854, 150

\bibitem[{{Alsina Ballester} {et~al.}(2019){Alsina Ballester}, {Belluzzi}, \&
  {Trujillo Bueno}}]{AlsinaBallester+19}
---. 2019, \apj, 880, 85

\bibitem[{{Belluzzi} {et~al.}(2012){Belluzzi}, {Trujillo Bueno}, \&
  {{\v{S}}t{\v{e}}p{\'a}n}}]{Belluzzi+12}
{Belluzzi}, L., {Trujillo Bueno}, J., \& {{\v{S}}t{\v{e}}p{\'a}n}, J. 2012,
  \apjl, 755, L2

\bibitem[{{Bommier}(1997)}]{Bommier97b}
{Bommier}, V. 1997, A\&A

\bibitem[{Clenshaw(1955)}]{Clenshaw55}
Clenshaw, C.~W. 1955, Mathematics of Computation, 9, 118

\bibitem[{{del Pino Alem{\'a}n} {et~al.}(2016){del Pino Alem{\'a}n}, {Casini},
  \& {Manso Sainz}}]{delPinoAleman+16}
{del Pino Alem{\'a}n}, T., {Casini}, R., \& {Manso Sainz}, R. 2016, \apjl, 830,
  L24

\bibitem[{{del Pino Alem{\'a}n} {et~al.}(2020){del Pino Alem{\'a}n}, {Trujillo
  Bueno}, {Casini}, \& {Manso Sainz}}]{delPinoAleman+20}
{del Pino Alem{\'a}n}, T., {Trujillo Bueno}, J., {Casini}, R., \& {Manso
  Sainz}, R. 2020, \apj, 891, 91

\bibitem[{Driscoll {et~al.}(2014)Driscoll, Hale, \& {L. N. Trefethen
  (editors)}}]{driscoll2014chebfun}
Driscoll, T.~A., Hale, N., \& {L. N. Trefethen (editors)}. 2014, Chebfun Guide
  (Pafnuty Publications, Oxford)

\bibitem[{{Fontenla} {et~al.}(1993){Fontenla}, {Avrett}, \&
  {Loeser}}]{Fontenla+93}
{Fontenla}, J.~M., {Avrett}, E.~H., \& {Loeser}, R. 1993, ApJ, 406, 319

\bibitem[{Golub \& Van~Loan(2013)}]{Golub13}
Golub, G.~H. \& Van~Loan, C.~F. 2013, Matrix Computations (4th ed., Johns
  Hopkins University Press)

\bibitem[{{Gouttebroze}(1986)}]{Gouttebroze86}
{Gouttebroze}, P. 1986, A\&A, 160, 195

\bibitem[{Hashemi \& Trefethen(2017)}]{Chebfun3}
Hashemi, B. \& Trefethen, L.~N. 2017, SIAM Journal on Scientific Computing, 39,
  C341

\bibitem[{{Hummer}(1962)}]{Hummer62}
{Hummer}, D.~G. 1962, MNRAS, 125, 21

\bibitem[{{Janett} {et~al.}(2017){Janett}, {Carlin}, {Steiner}, \&
  {Belluzzi}}]{Janett17a}
{Janett}, G., {Carlin}, E.~S., {Steiner}, O., \& {Belluzzi}, L. 2017, ApJ, 840,
  107

\bibitem[{{Janett} \& {Paganini}(2018)}]{JanettPaganini18}
{Janett}, G. \& {Paganini}, A. 2018, \apj, 857, 91

\bibitem[{{Kano} {et~al.}(2017){Kano}, {Trujillo Bueno}, {Winebarger},
  {Auch{\`e}re}, {Narukage}, {Ishikawa}, {Kobayashi}, {Bando}, {Katsukawa},
  {Kubo}, {Ishikawa}, {Giono}, {Hara}, {Suematsu}, {Shimizu}, {Sakao},
  {Tsuneta}, {Ichimoto}, {Goto}, {Belluzzi}, {{\v{S}}t{\v{e}}p{\'a}n}, {Asensio
  Ramos}, {Manso Sainz}, {Champey}, {Cirtain}, {De Pontieu}, {Casini}, \&
  {Carlsson}}]{Kano+17}
{Kano}, R., {Trujillo Bueno}, J., {Winebarger}, A., {et~al.} 2017, \apjl, 839,
  L10

\bibitem[{{Landi Degl'Innocenti} \& {Landolfi}(2004)}]{BLandiLandolfi04}
{Landi Degl'Innocenti}, E. \& {Landolfi}, M. 2004, Polarization in Spectral
  Lines (Klumer Academic Publishers)

\bibitem[{{Nagendra} \& {Sampoorna}(2011)}]{NagendraSampoorna11}
{Nagendra}, K.~N. \& {Sampoorna}, M. 2011, \aap, 535, A88

\bibitem[{{Paganini} \& {Hashemi}(2020)}]{Zenodo}
{Paganini}, A. \& {Hashemi}, B. 2020, Software to compute chebfun-based
  approximations of angle-averaged redistribution functions

\bibitem[{{Rees} \& {Saliba}(1982)}]{ReesSaliba82}
{Rees}, D.~E. \& {Saliba}, G.~J. 1982, A\&A, 115, 1

\bibitem[{{Sampoorna} {et~al.}(2019){Sampoorna}, {Nagendra}, {Sowmya},
  {Stenflo}, \& {Anusha}}]{Sampoorna+19}
{Sampoorna}, M., {Nagendra}, K.~N., {Sowmya}, K., {Stenflo}, J.~O., \&
  {Anusha}, L.~S. 2019, \apj, 883, 188

\bibitem[{{Sampoorna} {et~al.}(2017){Sampoorna}, {Nagendra}, \&
  {Stenflo}}]{Sampoorna+17}
{Sampoorna}, M., {Nagendra}, K.~N., \& {Stenflo}, J.~O. 2017, \apj, 844, 97

\bibitem[{{Trujillo Bueno}(2014)}]{TrujilloBueno14}
{Trujillo Bueno}, J. 2014, in ASP Conf. Ser., Vol. 489, Solar Polarization 7,
  ed. K.~N. {Nagendra}, J.~O. {Stenflo}, Z.~Q. {Qu}, \& M.~{Sampoorna}, 137

\bibitem[{{Trujillo Bueno} {et~al.}(2018){Trujillo Bueno},
  {{\v{S}}t{\v{e}}p{\'a}n}, {Belluzzi}, {Asensio Ramos}, {Manso Sainz}, {del
  Pino Alem{\'a}n}, {Casini}, {Ishikawa}, {Kano}, {Winebarger}, {Auch{\`e}re},
  {Narukage}, {Kobayashi}, {Bando}, {Katsukawa}, {Kubo}, {Ishikawa}, {Giono},
  {Hara}, {Suematsu}, {Shimizu}, {Sakao}, {Tsuneta}, {Ichimoto}, {Cirtain},
  {Champey}, {De Pontieu}, \& {Carlsson}}]{TrujilloBueno+18}
{Trujillo Bueno}, J., {{\v{S}}t{\v{e}}p{\'a}n}, J., {Belluzzi}, L., {et~al.}
  2018, \apjl, 866, L15

\bibitem[{{Uitenbroek}(1989)}]{Uitenbroek89}
{Uitenbroek}, H. 1989, A\&A, 216, 310

\end{thebibliography}
\bibliographystyle{apj}

\end{document}